\documentclass{article}

\newcommand{\FM}{{\varphi}}
\newcommand{\Fm}{{\phi}}
\newcommand{\imp}{\Rightarrow}
\newcommand{\lt}{r}
\newcommand{\scp}{\psi}
\newcommand{\X}{\odot}

\usepackage{xcolor}\usepackage{hyperref}

\title{\vspace{-2em}A Reply to ``On Salum’s Algorithm for X3SAT''}

\author{Latif Salum \\  latif.salum@gmail.com}

\date{}

\begin{document}

\maketitle

\begin{abstract}
This paper gives a reply to ``On Salum’s Algorithm for X3SAT'' \cite{paper}.
\end{abstract}

The authors confuse the definition of the scope $\scp(\lt_i)$, beyond the scope $\Fm'(\lt_i)$, and the formula $\FM = \scp \wedge \Fm$. Also, they claim that ``it is easy to see that X3SAT formulas with empty minterms are equivalent to 3CNF formulas (by simply replacing the $\X$ notational element with $\vee$)''. I did not understand what the authors meant here. The formula $(a \X b \X c) \wedge (b \X x \X y) \wedge (c \X x \X \overline{y})$ is totally different from the formula $(a \vee b \vee c) \wedge (b \vee x \vee y) \wedge (c \vee x \vee \overline{y})$.

In particular, the authors find $\color{red}\Fm = (a \wedge \overline{b} \wedge \overline{c}) \wedge (x \X y) \wedge (x \X \bar{y})$, after \texttt{Scope}~$\!(a, \Fm)$ runs. They  next run \texttt{Scope}~$\!(\overline{a}, \Fm)$ over $\color{red}\Fm$, and detect the conflict with $a$ already in the minterm. The authors also run \texttt{Scope}~$\!(x, \Fm)$ over $\color{red}\Fm$, and find $\color{red}\Fm = (a \wedge \overline{b} \wedge \overline{c} \wedge x \wedge \bar{y} \wedge y)$. These are totally wrong.

The correct way of the executions is as follows.

$\FM = \scp \wedge \Fm$. $\scp$ is empty and $\color{blue}\Fm = (a \X b \X c) \wedge (b \X x \X y) \wedge (c \X x \X \overline{y})$.

\texttt{Scope}~$\!(a, \Fm)$: $\color{orange}\Fm(a) := a \wedge \Fm$. Then, $\Fm(a) = \scp(a) \wedge \Fm'(a)$, as \texttt{Scope}~$\!(a, \Fm)$ terminates. $\color{blue}\scp(a) = a \wedge \overline{b} \wedge \overline{c}$ and $\color{blue}\Fm'(a) = (x \X y) \wedge (x \X \bar{y})$.

\texttt{Scope}~$\!(\overline{a}, \Fm)$: $\Fm(\overline{a}) := \overline{a} \wedge \Fm$. $\Fm(\overline{a}) = \scp(\overline{a}) \wedge \Fm'(\overline{a})$. $\color{blue}\scp(\overline{a}) = \overline{a}$ and $\color{blue}\Fm'(\overline{a}) = (b \X c) \wedge (b \X x \X y) \wedge (c \X x \X \overline{y})$.

\texttt{Scope}~$\!(x, \Fm)$: $\Fm(x) := x \wedge \Fm$. $\scp(x) \imp \overline{b} \wedge x \wedge \overline{y} \wedge \overline{c} \wedge y$. Thus, $x$ is incompatible, because $\scp(x) \imp  \overline{y} \wedge y$. As a result, $\overline{x}$ is necessary. That is, $\FM \gets  \overline{x} \wedge \Fm$.

$\FM = \scp \wedge \Fm$, where $\scp = \overline{x}$ and $\color{blue}\Fm = (a \X b \X c) \wedge (b \X y) \wedge (c \X \overline{y})$.

\texttt{Scope}~$\!(a, \Fm)$: $\color{orange}\Fm(a) := a \wedge \Fm$.  $\scp(a) \imp a \wedge \overline{b} \wedge \overline{c} \wedge y \wedge \overline{y}$. Thus, $\overline{a}$ is necessary.

$\FM \gets \scp \wedge \Fm$, where $\scp = \overline{x} \wedge \overline{a}$ and $\color{blue}\Fm = (b \X c) \wedge (b \X y) \wedge (c \X \overline{y})$.

\texttt{Scope}~$\!(b, \Fm)$: $\Fm(b) := b \wedge \Fm$.  $\scp(b) = b \wedge \overline{c} \wedge \overline{y}$ and $\Fm'(b)$ is empty.

\texttt{Scope}~$\!(\overline{b}, \Fm)$: $\Fm(\overline{b}) := \overline{b} \wedge \Fm$.  $\scp(\overline{b}) = \overline{b} \wedge c \wedge y$ and $\Fm'(\overline{b})$ is empty.

Likewise, $c, \overline{c}, y, \overline{y}$ are decided to be compatible. Then,  the current formula  becomes $\FM = \overline{x} \wedge \overline{a} \wedge \color{blue} (b \X c) \wedge (b \X y) \wedge (c \X \overline{y})$, which is satisfiable. Note that $\overline{x} \wedge \overline{a}$ is fixed in any satisfying assignment. Next, a satisfying assignment is constructed over $\color{blue}\Fm = (b \X c) \wedge (b \X y) \wedge (c \X \overline{y})$. See also this short presentation \cite{salum}.


\begin{thebibliography}{2}

\bibitem{paper}
Arian Nadjimzadah and David E. Narváez (2021). On Salum's Algorithm for $\mathrm{X3SAT}$. arXiv:2104.02886

\bibitem{salum}
Latif Salum (2024). On Syntax and Semantics of PL.  \url{https://doi.org/10.13140/RG.2.2.14328.05120}

\end{thebibliography}
\end{document}